\documentclass[twocolumn,showpacs,preprintnumbers,amsmath,amssymb]{revtex4}
\usepackage{graphicx}% Include figure files
\usepackage{dcolumn}% Align table columns on decimal point
\usepackage{bm}% bold math

\begin{document}
\title{Study on the properties of the coupling constants of $J/\psi \rightarrow VP$ decays}
\author{Jinshu Huang}
\email{drjshuang@hotmail.com} \affiliation{College of Physics $\&$
Electronic Engineering, Nanyang Normal University, Nanyang 473061,
People's Republic of China; \\ College of Physics $\&$ Electronic
Engineering, Henan Normal University, Xinxiang 453007, People's
Republic of China}
\author{Junfeng Sun} \email{sunjunfeng@htu.cn}
\author{Gongru Lu} \email{lugongru@sina.com}
\affiliation{College of Physics $\&$ Electronic
Engineering, Henan Normal University, Xinxiang 453007, People's
Republic of China}
\author{Haibo Li}
\email{hbli@ihep.ac.cn} \affiliation{Institute of High Energy Physics, Chinese Academy of Sciences, Beijing 100049,
People's Republic of China}

\date{\today}

\begin{abstract}
Basing on the branching fractions of $J/\psi \to VP$ from different
experiments,  we investigate on the properties of the coupling constants of $J/\psi
\rightarrow VP$ decays with a model-dependent approach. We find that the octet coupling constant, $g_8$, of
strong interaction is about twice larger than that of the
singlet coupling constant $g_1$; the electromagnetic breaking
parameters $g^i_E$ are larger than the mass breaking parameters $g^i_M$,
moreover, the three parameters of electromagnetic effect are about
equal, but the three parameters of mass effect are obviously
different and their uncertainties are also large; and the phase angle
between strong and electromagnetic interaction is in the range of $70^{\circ} \sim 80^{\circ}$. It deepens our understanding
of the coupling constant of $J/\psi \rightarrow VP$ decays.
\end{abstract}


\maketitle

\section{\label{sec:level1}Introduction}

The $J/\psi$ decays play an important role in the understanding of low-energy hadron dynamics.  A particular
advantage of these decays is that the initial state is a very good
approximation of a flavor-$SU(3)$ singlet. In order to estimate
that $SU(3)$ is an approximate symmetry of low-energy hadron
interactions, we need study in detail of the information on the
final-state particles of $J/\psi$ decay.

The implication of $SU(3)$ symmetry to $J/\psi$ decays into mesons
has been studied \cite{pdg2012}.  $J/\psi$ can decay into a vector
and a pseudoscalar induced by three-gluon annihilation and
electromagnetic processes. However, these branching ratios are only
of the order $10^{-3}$ since the hadronic decays of $J/\psi$ are
suppressed by the Okubo-Zweig-Iizuka (OZI) rule \cite{haber1985}.

Based on the $SU(3)$ symmetry, three quarks may make
up an octet and one singlet.  Because the mass of $s$ quark is
larger than those of $u$ and $d$ quarks and the mixing in the mesons
exists, it will cause $SU(3)$ breaking and further create a physical
nonet. The octet and singlet are the eigenstates of $SU(3)$ group, however, what we can
obverse in experiment are not $SU(3)$ eigenstates themselves, but
their mixing. For example, for the pseudoscalar mesons, if
pseudoscalar glueballs and radially excited states are ignored and
only quark states are considered, then the physical states $\eta$
and $\eta'$ are related to $\eta^8$ and $\eta^0$, via the usual
mixing formulas
 \begin{eqnarray}
 \left( \begin{array}{c}  {\eta} \\ {\eta'}  \end{array} \right) \,=\,
 \left( \begin{array}{cc}  {\cos}{\theta}_{P} & -{\sin}{\theta}_{P} \\  {\sin}{\theta}_{P} &  {\cos}{\theta}_{P}  \end{array} \right)
 \left( \begin{array}{c}  \eta^{8} \\ \eta^{0}  \end{array} \right),
\end{eqnarray}
 where $\theta_P$ is the mixing angle between $\eta^8$ and $\eta^0$.  Similarly, for the vector mesons $\phi$ and $\omega$, we have
 \begin{eqnarray}
 \left( \begin{array}{c}  {\phi} \\ {\omega}  \end{array} \right) \,=\,
 \left( \begin{array}{cc}  {\cos}{\theta}_{V} & -{\sin}{\theta}_{V} \\  {\sin}{\theta}_{V} &  {\cos}{\theta}_{V}  \end{array} \right)
 \left( \begin{array}{c}  \omega^{8} \\ \omega^{0}  \end{array} \right),
\end{eqnarray}
where $\theta_V$ is the mixing angle between $\omega^8$ and $\omega^0$.

For vector mesons, the mixing between $\omega^8$ and $\omega^0$ is
basically an idea mixing,  ${\sin}{\theta}_{V}=\sqrt{1/3}$,
${\cos}{\theta}_{V}=\sqrt{2/3}$, i. e. $\theta_V \approx
35.3^{\circ}$, so in this case we have
 \begin{eqnarray}
\omega=\sqrt{\frac{1}{2}}|u\bar{u}+d\bar{d}>,\ \phi=|s\bar{s}>.
  \end{eqnarray}
However, for pseudoscalar mesons, it always brings people great
interesting  to discuss the mixing between $\eta^8$ and $\eta^0$
\cite{thomas2007}-\cite{donoto2012}.

Combining the mixings in vector mesons and pseudoscalar mesons and
considering  the various mass effects and electromagnetic effects,
in this paper we shall analyse the properties of the coupling constants of $J/\psi
\to VP$ decay, which is very important to comprehend the breaking of
$SU(3)$ flavor symmetry.

\section{Phenomenological study on the decays $J/\psi \to VP$}

\subsection{Effective Lagrangian of $J/\psi \rightarrow VP$}

For the two-body decays $J/\psi \rightarrow H_1 H_2$, where $H_1$
and $H_2$ denote mesons,  the most general interaction term in an
$SU(3)$ invariant is \cite{haber1985}
\begin{equation}
L_{\rm int}=\psi (g_8 O^{a}_1 O^{a}_2 +g_1 S_1 S_2),
\end{equation}
where $O$ and $S$ denote an octet and a singlet, respectively, and a sum over $a=1, 2, \cdots, 8$ is implied.

The $SU(3)$ symmetry in the nonet of pseudoscalar meson is not
strict, two types of $SU(3)$ breaking  should be considered. The first one
is induced by the different mass of quark, and the other one is the
different charge of quark.

In our theoretical calculation, $m_u=m_d$ is usually assumed, but
$m_s \neq m_d$,  this difference of mass will cause the breaking
effect on $SU(3)$ symmetry
\begin{equation}
m_d(\bar{u}u+\bar{d}d)+m_s \bar{s}s=m_0 \bar{q}q +\frac{1}{\sqrt{3}}(m_d-m_s) \bar{q} \lambda_8 q,
\end{equation}
where $q=(u,\ d,\ s)$, $m_0=\frac{1}{3}(2m_d+m_s)$ is the average
quark mass,  and $\lambda_8$ is the eighth one of Gell-Mann matrices.
The last term in Eq. (5) is the mass effects of violating $SU(3)$
invariance. We need introduce a new spurion $M^a=\delta^{a8}$ to
describe this $SU(3)$ breaking term.

The electromagnetic effects violate $SU(3)$ symmetry since the photon coupling to quarks is proportional to the electric charge
\begin{equation}
\frac{2}{3}\bar{u} \gamma_{\mu}u-\frac{1}{3}\bar{d}\gamma_{\mu}d-\frac{1}{3} \bar{s}\gamma_{\mu}s
=\frac{1}{2} \bar{q} \gamma_{\mu} (\lambda_3+\frac{1}{\sqrt{3}}\lambda_8)q,
\end{equation}
where $\lambda_3$ and $\lambda_8$ denote the third and eighth of
Gell-Mann matrices, respectively.  It follows from Eq. (6) that the
electromagnetic breaking can be simulated by a spurion
$E=\delta^{a3}+\frac{1}{\sqrt{3}} \delta ^{a8}$.

After the above two effects of $SU(3)$ breaking are considered, the effective Lagrangian of $J/\psi \rightarrow H_1 H_2$ process can be written \cite{haber1985}
 \begin{eqnarray}
 && {\cal L}_{\rm eff} = {\psi} \bigg \{ g_{8}O_{1}^{a}O_{2}^{a}+g_{1}S_{1}S_{2} +g_{S}d_{abc}O_{1}^{a}O_{2}^{b}O_{3}^{c} \nonumber \\
 && +g_{A}f_{abc}O_{1}^{a}O_{2}^{b}O_{3}^{c} +\sqrt{\frac{2}{3}} \bigg [ C_{123}O_{1}^{a}O_{2}^{a}S_{3}   \nonumber \\
 && +C_{132}O_{1}^{a}O_{3}^{a}S_{2}+C_{231}O_{2}^{a}O_{3}^{a}S_{1}
 +f\,S_{1}S_{2}S_{3}\bigg] \bigg \}.
  \end{eqnarray}

If the final states of $J/\psi \rightarrow H_1 H_2$ decays are a
vector and a pseudoscalar,  the expression of the above effective
Lagrangian can be further written as
\begin{eqnarray}
 && {\cal L}_{\rm eff}= \psi \bigg \{ g_{8}P_{1}^{a}V_{2}^{a}+g_{1}P_{0}V_{0} +g_{M}^{88}d_{abc}O_{1}^{a}O_{2}^{b}M_{3}^{c}
  \nonumber \\
 && + \sqrt{\frac{2}{3}}\bigg [ g_{M}^{81}O_{1}^{a}M_{3}^{a}S_{2} +g_{M}^{18}O_{2}^{a}M_{3}^{a}S_{1} \bigg ]  +g_{E}^{88}d_{abc}O_{1}^{a}  \nonumber \\
 && \times O_{2}^{b}E_{3}^{c} + \sqrt{\frac{2}{3}} \bigg [ g_{E}^{81}O_{1}^{a}E_{3}^{a}S_{2}
 +g_{E}^{18}O_{2}^{a}E_{3}^{a}S_{1} \bigg ] \bigg \},
 \end{eqnarray}
\noindent in which  $g_8$ and $g_1$ denote the coupling constants of
octet and singlet,  $g^i_M$ and $g^i_E$ are the coupling constants
of the mass breaking term and electromagnetic breaking term,
respectively, $f_{abc}$ and $d_{abc}$ coefficients are the
antisymmetrical and symmetrical structure constants of $SU(3)$ group.

\subsection{Decay amplitude and width of $J/\psi \to VP$ decays}

The physical particles in correspondence with the pseudoscalar and vector mesons are
\begin{eqnarray}
 P_{1}&=& \frac{1}{\sqrt{2}}({\pi}^{+}+{\pi}^{-}), \nonumber \\
 P_{2}&=& \frac{i}{\sqrt{2}}({\pi}^{+}-{\pi}^{-}), \nonumber \\
 P_{3}&=& {\pi}^{0}, \nonumber \\
 P_{4}&=& \frac{1}{\sqrt{2}}(K^{+}+K^{-}), \nonumber \\
 P_{5}&=& \frac{i}{\sqrt{2}}(K^{+}-K^{-}), \nonumber \\
 P_{6}&=& \frac{1}{\sqrt{2}}(K^{0}+\overline{K}^{0})\,=\,K_{S}^{0}, \nonumber \\
 P_{7}&=& \frac{i}{\sqrt{2}}(K^{0}-\overline{K}^{0})\,=\,iK_{L}^{0}, \nonumber  \\
 P_{8}&=& {\eta}_{8}^{0}\,=\, \eta \cos \theta_P +\eta' \sin \theta_P, \nonumber \\
 P_{0}&=&{\eta}_{1}^{0}\,=\, -\eta \sin \theta_P +\eta' \cos\theta_P,
\end{eqnarray}
and
\begin{eqnarray}
V_{1}&=& \frac{1}{\sqrt{2}}({\rho}^{+}+{\rho}^{-}), \nonumber \\
V_{2}&=&\frac{i}{\sqrt{2}}({\rho}^{+}-{\rho}^{-}), \nonumber \\
V_{3} &=& {\rho}^{0}, \nonumber \\
V_{4} &=& \frac{1}{\sqrt{2}}(K^{{\ast}+}+K^{{\ast}-}), \nonumber \\
V_{5} &=& \frac{i}{\sqrt{2}}(K^{{\ast}+}-K^{{\ast}-}), \nonumber \\
V_{6} &=& \frac{1}{\sqrt{2}}(K^{{\ast}0}+\overline{K}^{{\ast}0}), \nonumber \\
V_{7}&=& \frac{i}{\sqrt{2}}(K^{{\ast}0}-\overline{K}^{{\ast}0}), \nonumber \\
V_{8} &=& \phi \cos \theta_V +\omega \sin \theta_V, \nonumber \\
V_{0} &=&  -\phi \sin \theta_V +\omega \cos\theta_V,
\end{eqnarray}
\noindent then we calculate in detail on the effective Lagrangian of  $J/\psi \to VP$ decays
\begin{eqnarray}
  P_{1}^{a}V_{2}^{a} &=&  P^{1}V^{1}+P^{2}V^{2}+P^{3}V^{3}+P^{4}V^{4}
  +P^{5}V^{5}\nonumber \\ && +P^{6}V^{6}+P^{7}V^{7}+P^{8}V^{8} \nonumber \\
  &=& {\pi}^{+}{\rho}^{-}+{\pi}^{-}{\rho}^{+}+{\pi}^{0}{\rho}^{0}+ K^{+}K^{{\ast}-}\nonumber \\
  && +K^{-}K^{{\ast}+}+K^{0}\overline{K}^{{\ast}0}+
  \overline{K}^{0}K^{{\ast}0} \nonumber \\
  && + {\eta}{\omega}{\cos}{\theta}_{P}{\sin}{\theta}_{V}+ {\eta}{\phi}{\cos}{\theta}_{P}
  {\cos}{\theta}_{V} \nonumber \\ && +  {\eta}^{\prime}{\omega}{\sin}{\theta}_{P}{\sin}{\theta}_{V}+ {\eta}^{\prime}{\phi}{\sin}{\theta}_{P}{\cos}{\theta}_{V},
 \end{eqnarray}
\begin{eqnarray}
  S_{1}S_{2} &=& P^{0}V^{0} = -{\eta}{\omega} {\sin}{\theta}_{P}{\cos}{\theta}_{V} +{\eta}{\phi}   {\sin}{\theta}_{P}{\sin}{\theta}_{V}
  \nonumber \\
  && +{\eta}^{\prime}{\omega} {\cos}{\theta}_{P}{\cos}{\theta}_{V} -{\eta}^{\prime}{\phi}   {\cos}{\theta}_{P}{\sin}{\theta}_{V},
 \end{eqnarray}
\begin{eqnarray}
 d_{abc}O_{1}^{a}O_{2}^{b}M_{3}^{c}&=& d_{ab8}P^{a}V^{b} =
 d_{118}P^{1}V^{1}+d_{228}P^{2}V^{2} \nonumber \\
  && +d_{338}P^{3}V^{3}+d_{448}P^{4}V^{4}+d_{558}P^{5}V^{5}\nonumber \\
 &&  + d_{668}P^{6}V^{6}+d_{778}P^{7}V^{7}+d_{888}P^{8}V^{8} \nonumber \\
  &=&  \frac{1}{\sqrt{3}} \Big[ {\pi}^{+}{\rho}^{-}+{\pi}^{-}{\rho}^{+} +{\pi}^{0}{\rho}^{0} \Big] \nonumber \\
  && \,+\,
 \frac{-1}{2\sqrt{3}} \Big[ K^{+}K^{{\ast}-}+K^{-}K^{{\ast}+}+K^{0}\overline{K}^{{\ast}0}\nonumber \\
  &&  +\overline{K}^{0}K^{{\ast}0} \Big] + \frac{-1}{\sqrt{3}} \Big[ {\eta}{\omega}{\cos}{\theta}_{P}{\sin}{\theta}_{V} \nonumber \\
 && + {\eta}{\phi}{\cos}{\theta}_{P}{\cos}{\theta}_{V} + {\eta}^{\prime}{\omega}{\sin}{\theta}_{P}{\sin}{\theta}_{V}
 \nonumber \\
  && + {\eta}^{\prime}{\phi}{\sin}{\theta}_{P}{\cos}{\theta}_{V} \Big],
\end{eqnarray}
 \begin{eqnarray}
 O_{1}^{a}M_{3}^{a}S_{2}&=& P^{8}V^{0} = {\eta}{\omega}  {\cos}{\theta}_{P}{\cos}{\theta}_{V} -{\eta}{\phi}    {\cos}{\theta}_{P}{\sin}{\theta}_{V}
  \nonumber \\
  && +{\eta}^{\prime}{\omega} {\sin}{\theta}_{P}{\cos}{\theta}_{V} -{\eta}^{\prime}{\phi}   {\sin}{\theta}_{P}{\sin}{\theta}_{V},
 \end{eqnarray}
 \begin{eqnarray}
 O_{2}^{a}M_{3}^{a}S_{1} &=& P^{0}V^{8} = -{\eta}{\omega}  {\sin}{\theta}_{P}{\sin}{\theta}_{V} -{\eta}{\phi}    {\sin}{\theta}_{P}{\cos}{\theta}_{V}
 \nonumber \\
  &&  +{\eta}^{\prime}{\omega} {\cos}{\theta}_{P}{\sin}{\theta}_{V} +{\eta}^{\prime}{\phi}   {\cos}{\theta}_{P}{\cos}{\theta}_{V},
 \end{eqnarray}
  \begin{eqnarray}
  d_{ab3}P^{a}V^{b}&=& d_{383}P^{3}V^{8}+d_{833}P^{8}V^{3}
  +d_{443}P^{4}V^{4}\nonumber \\
  && +d_{553}P^{5}V^{5}+d_{663}P^{6}V^{6}+d_{773}P^{7}V^{7} \nonumber \\
 &=& {\pi}^{0}{\omega}\sqrt{\frac{1}{3}}{\sin}{\theta}_{V}+{\pi}^{0}{\phi}\sqrt{\frac{1}{3}}{\cos}{\theta}_{V}
    \nonumber \\
  && +{\eta}{\rho}^{0}\sqrt{\frac{1}{3}}{\cos}{\theta}_{P}+{\eta}^{\prime}{\rho}^{0}\sqrt{\frac{1}{3}}{\sin}{\theta}_{P}
    \nonumber \\
 && + \frac{1}{2}K^{+}K^{{\ast}-} +\frac{1}{2}K^{-}K^{{\ast}+}\nonumber \\
 && -  \frac{1}{2}K^{0}\overline{K}^{{\ast}0} - \frac{1}{2}\overline{K}^{0}K^{{\ast}0},
 \end{eqnarray}
  \begin{eqnarray}
 d_{abc}O_{1}^{a}O_{2}^{b}E_{3}^{c}= d_{ab3}P^{a}V^{b}+\frac{1}{\sqrt{3}}d_{ab8}P^{a}V^{b},
 \end{eqnarray}
\begin{eqnarray}
 O_{1}^{a}E_{3}^{a}S_{2}=P^{3}V^{0}+\frac{1}{\sqrt{3}}P^{8}V^{0},
  \end{eqnarray}
  \begin{eqnarray}
P^{3}V^{0} &=&   {\pi}^{0}\left({\omega}{\cos}{\theta}_{V}-{\phi}{\sin}{\theta}_{V}\right) \nonumber \\
&=&   \sqrt{\frac{2}{3}}{\pi}^{0}{\omega}-\sqrt{\frac{1}{3}}{\pi}^{0}{\phi},
   \end{eqnarray}
  \begin{eqnarray}
 O_{2}^{a}E_{3}^{a}S_{1}= V^{3}P^{0}+\frac{1}{\sqrt{3}}V^{8}P^{0},
  \end{eqnarray}
  \begin{eqnarray}
V^{3}P^{0}  &=&  {\rho}^{0}\left(-{\eta}{\sin}{\theta}_{P}+{\eta}^{\prime}{\cos}{\theta}_{P}\right) \nonumber \\
&=&   -{\eta}{\rho}^{0}{\sin}{\theta}_{P}+{\eta}^{\prime}{\rho}^{0}{\cos}{\theta}_{P},
 \end{eqnarray}
therefore we obtain the coupling constants and their corresponding
strengths of $J/\psi \to PV$ decays given in  Table 1. It should be noted
that our results are consistent with Ref. \cite{haber1985}, but we
distinguish in detail the different parameters of mass effect and
electromagnetic effect.

\begin{table*}
\begin{center}
{\small Table\ 1\ \ The coupling constants and their corresponding strengths of $J/\psi \to PV$.}
\end{center}

 {\scriptsize 
 %%\footnotesize
 %%\tiny 
 \[ \!\!\!\!\!\!\!
  \begin{array}{l}
  \begin{array}{l||c|c||c|c|c||c|c|c} \hline
  \multicolumn{1}{c||}{\rm Decay\ modes} &
  \multicolumn{8}{c}{\rm Coupling\ constants\ and\ thier\ corresponding\ strengths} \\ \cline{2-9}
  J/{\psi}{\to}PV & g_{8} & g_{1} & g_{M}^{88} & g_{M}^{81}
  & g_{M}^{18} & g_{E}^{88} & g_{E}^{81} & g_{E}^{18}  \\ \hline
 {\pi}^{+}{\rho}^{-} & 1 & 0 & \frac{1}{\sqrt{3}} & 0 & 0 & \frac{1}{3} & 0 & 0 \\
 {\pi}^{-}{\rho}^{+} & 1 & 0 & \frac{1}{\sqrt{3}} & 0 & 0 & \frac{1}{3} & 0 & 0 \\
 {\pi}^{0}{\rho}^{0} & 1 & 0 & \frac{1}{\sqrt{3}} & 0 & 0 & \frac{1}{3} & 0 & 0 \\
 K^{+}K^{{\ast}-}    & 1 & 0 & \frac{-1}{2\sqrt{3}} & 0 & 0 & \frac{1}{3} & 0 & 0 \\
 K^{-}K^{{\ast}+}    & 1 & 0 & \frac{-1}{2\sqrt{3}} & 0 & 0 & \frac{1}{3} & 0 & 0 \\
 K^{0}\overline{K}^{{\ast}0} & 1 & 0 & \frac{-1}{2\sqrt{3}} & 0 & 0 & \frac{-2}{3} & 0 & 0 \\
 \overline{K}^{0}K^{{\ast}0} & 1 & 0 & \frac{-1}{2\sqrt{3}} & 0 & 0 & \frac{-2}{3} & 0 & 0 \\ \hline
 {\pi}^{0}{\omega}         & 0 & 0 & 0 & 0 & 0 & \!\!  \sqrt{\frac{1}{3}}{\sin}{\theta}_{V} & \!\! \sqrt{\frac{2}{3}}{\cos}{\theta}_{V} & 0 \\
 {\pi}^{0}{\phi}           & 0 & 0 & 0 & 0 & 0 & \!\! -\sqrt{\frac{1}{3}}{\cos}{\theta}_{V} & \!\! \sqrt{\frac{2}{3}}{\sin}{\theta}_{V} & 0 \\
 {\eta}{\rho}^{0}          & 0 & 0 & 0 & 0 & 0 & \!\!  \sqrt{\frac{1}{3}}{\cos}{\theta}_{P} & 0 & \!\! -\sqrt{\frac{2}{3}}{\sin}{\theta}_{P} \\
 {\eta}^{\prime}{\rho}^{0} & 0 & 0 & 0 & 0 & 0 & \!\!  \sqrt{\frac{1}{3}}{\sin}{\theta}_{P} & 0 & \!\!  \sqrt{\frac{2}{3}}{\cos}{\theta}_{P} \\ \hline
 {\eta}{\omega}    &  \!\! {\cos}{\theta}_{P}{\sin}{\theta}_{V}
                   &  \!\! -{\sin}{\theta}_{P}{\cos}{\theta}_{V}
                   &  \!\! \frac{-1}{\sqrt{3}}{\cos}{\theta}_{P}{\sin}{\theta}_{V}
                   &  \!\! \sqrt{\frac{2}{3}}{\cos}{\theta}_{P}{\cos}{\theta}_{V}
                   &  \!\! -\sqrt{\frac{2}{3}}{\sin}{\theta}_{P}{\cos}{\theta}_{V}
                   &  \!\! \frac{1}{\sqrt{3}}\hbox{4th colu.}
                   &  \!\! \frac{1}{\sqrt{3}}\hbox{5th colu.}
                   &  \!\! \frac{1}{\sqrt{3}}\hbox{6th colu.}  \\ \hline
 {\eta}{\phi}      &  \!\! {\cos}{\theta}_{P}{\cos}{\theta}_{V}
                   &  \!\!  {\sin}{\theta}_{P}{\sin}{\theta}_{V}
                   &  \!\! -\frac{1}{\sqrt{3}}{\cos}{\theta}_{P}{\cos}{\theta}_{V}
                   &  \!\! -\sqrt{\frac{2}{3}}{\cos}{\theta}_{P}{\sin}{\theta}_{V}
                   &  \!\! -\sqrt{\frac{2}{3}}{\sin}{\theta}_{P}{\sin}{\theta}_{V}
                   &  \!\! \frac{1}{\sqrt{3}}\hbox{4th colu.}
                   &  \!\! \frac{1}{\sqrt{3}}\hbox{5th colu.}
                   &  \!\!  \frac{1}{\sqrt{3}}\hbox{6th colu.}  \\ \hline
 {\eta}^{\prime}{\omega} &   \!\! {\sin}{\theta}_{P}{\sin}{\theta}_{V}
                         &  \!\! {\cos}{\theta}_{P}{\cos}{\theta}_{V}
                         &  \!\! \frac{-1}{\sqrt{3}}{\sin}{\theta}_{P}{\sin}{\theta}_{V}
                         &  \!\! \sqrt{\frac{2}{3}}{\sin}{\theta}_{P}{\cos}{\theta}_{V}
                         &  \!\! \sqrt{\frac{2}{3}}{\cos}{\theta}_{P}{\cos}{\theta}_{V}
                         &  \!\! \frac{1}{\sqrt{3}}\hbox{4th colu.}
                         &  \!\! \frac{1}{\sqrt{3}}\hbox{5th colu.}
                         &  \!\! \frac{1}{\sqrt{3}}\hbox{6th colu.} \\ \hline
 {\eta}^{\prime}{\phi}   &  \!\! {\sin}{\theta}_{P}{\cos}{\theta}_{V}
                         &  \!\! -{\cos}{\theta}_{P}{\sin}{\theta}_{V}
                         &  \!\! -\frac{1}{\sqrt{3}}{\sin}{\theta}_{P}{\cos}{\theta}_{V}
                         &  \!\! -\sqrt{\frac{2}{3}}{\sin}{\theta}_{P}{\sin}{\theta}_{V}
                         &  \!\! \sqrt{\frac{2}{3}}{\cos}{\theta}_{P}{\sin}{\theta}_{V}
                         &  \!\! \frac{1}{\sqrt{3}}\hbox{4th colu.}
                         &  \!\! \frac{1}{\sqrt{3}}\hbox{5th colu.}
                         &  \!\! \frac{1}{\sqrt{3}}\hbox{6th colu.} \\ \hline
 \end{array} \\ ~\\
\end{array}
 \] }
\end{table*}

 The amplitude of $J/{\psi} \to VP$ decays is
 \begin{equation}
  {\cal M}\,=\,\frac{g_{{\psi}VP}}{m_{\psi}} {\varepsilon}_{{\mu}{\nu}{\rho}{\sigma}}p_{\psi}^{\mu}
  {\varepsilon}_{\psi}^{\nu}p_{V}^{\rho}{\varepsilon}_{V}^{{\ast}{\sigma}},
 \end{equation}
where $g_{{\psi}VP}$ is the coupling constant and its various expressions are tabulated in Table 1, so we have
 \begin{eqnarray}
  && \overline{{\vert}{\cal M}{\vert}^{2}} =
  \frac{1}{2s_{\psi}+1}\sum\limits_{s_{\psi}}\sum\limits_{s_{V}}{\cal M}{\cal M}^{\dag}
  \nonumber \\ &&=
  \frac{2{\vert}g_{{\psi}VP}{\vert}^{2}}{3m_{\psi}^{2}}
  \left\{\left(\frac{m_{\psi}^{2}+m_{V}^{2}-m_{P}^{2}}{2}\right)^{2}
  -m_{\psi}^{2}m_{V}^{2}\right\}
   \nonumber \\ &&=
  \frac{2}{3}{\vert}g_{{\psi}VP}{\vert}^{2}p^{2},
 \end{eqnarray}
 in which $\vec{p}$ is the momenta of vector meson in the c.m. frame, and can be denoted as
  \begin{equation}
p=|\vec{p}|=\frac{\sqrt{\left[m_{\psi}^{2}-(m_{V}+m_{P})^{2}\right] \left[m_{\psi}^{2}-(m_{V}-m_{P})^{2}\right]}}{2m_{\psi}} .
 \end{equation}

The differential decay rate for $J/{\psi} \to VP$ can be written as
 \begin{equation}
 {\rm d}{\Gamma}\,=\,\frac{1}{32{\pi}^{2}} \frac{p}{m_{\psi}^{2}}\overline{{\vert}{\cal M}{\vert}^{2}} {\rm d}{\Omega},
 \end{equation}
so we have the total decay width
 \begin{equation}
 {\Gamma}\,=\,  {\int}\frac{1}{32{\pi}^{2}} \frac{p}{m_{\psi}^{2}} \overline{{\vert}{\cal M}{\vert}^{2}} {\rm d}{\Omega}
 \,=\,  \frac{1}{3} \frac{p^{3}}{m_{\psi}^{2}} \frac{{\vert}g_{{\psi}VP}{\vert}^{2}}{4{\pi}} .
 \end{equation}

\section{Numerical analysis and discussions}

In Table 1, we give out the free parameters involved in the decays
$J/\psi \to PV$. In these  parameters, $g^{88}_M$, $g^{81}_M$ and
$g^{18}_M$ denote the $SU(3)$ mass breaking terms induced by the
different quark mass, $g^{88}_E$, $g^{81}_E$ and $g^{18}_E$ denote
the $SU(3)$ electromagnetic terms induced by the different quark
electric charge. There is a phase angle, noted by $\delta_E$,
between electromagnetic and strong interaction. We can take the
coupling constants of mass effect as real, and denote the
coupling constants of electromagnetic effect with $g^i_E=|g^i_E|
e^{\delta_E}$. Therefore there are eleven free parameters in the
decays $J/\psi \to PV$, they are $g_8$, $g_1$, $g^{88}_M$,
$g^{81}_M$, $g^{18}_M$, $|g^{88}_E|$, $|g^{81}_E|$, $|g^{18}_E|$,
$\delta_E$, $\theta_P$ and $\theta_V$. For the sake of simplicity,
we shall take the following four cases to analyze and discuss the
properties of these parameters:

Case I: Assuming $g^{88}_M=g^{81}_M=g^{18}_M=g_M$,
$g^{88}_E=g^{81}_E=g^{18}_E=g_E$,  and taking the mixing between
$\phi$ and $\omega$ regard as idea mixing, i.e., $\theta_V \approx
35.3^{\circ}$. In this case, there are only six parameters, they are
$g_8$, $g_1$, $g_M$, $|g_E|$, $\delta_E$ and $\theta_P$.

Case II: Assuming $g^{88}_M=g^{81}_M=g^{18}_M=g_M$,
$g^{88}_E=g^{81}_E=g^{18}_E=g_E$,  but taking the mixing angle
$\theta_V$ regard as free. In this case, the parameters have seven which 
are $g_8$, $g_1$, $g_M$, $|g_E|$, $\delta_E$, $\theta_P$,
$\theta_V$ respectively.

Case III: Assuming the three parameters of mass effects aren't
equal,  and the three parameters of electromagnetic effects aren't
equal, either, they are all free parameters, but taking $\theta_V
\approx 35.3^{\circ}$, then we shall have 10 parameters: $g_8$,
$g_1$, $g^{88}_M$, $g^{81}_M$, $g^{18}_M$, $|g^{88}_E|$,
$|g^{81}_E|$, $|g^{18}_E|$, $\delta_E$, and $\theta_P$.

Case IV: Assuming the three parameters of mass effects aren't equal,
and the three parameters of electromagnetic effects aren't equal,
either, further taking the mixing angle $\theta_V$ as free in
the same, then all of the eleven parameters are free.

For the latter two cases, we have to request more experimental
information to analyze them due to too many parameters.

The experimental results for the decays $J/\psi \to PV$ are mainly from Mark-III \cite{mark31988}, DM2 
\cite{dm21990} and BES \cite{bes2004}-\cite{bes2006}, which are shown in Table 2. The last column in this table is the latest world average
in 2012 \cite{pdg2012}. To clarify the results obtained from
different data set, we divided them into three subsections to
investigate the properties of the coupling constants of the decays
$J/\psi \to PV$.

\subsection{Analysis of $J/\psi \to VP$ from MarkIII, DM2 and PDG2012 data in Case I}

First, we consider the simplest case, i.e., Case I. In Ref.
\cite{wei2010}, the pseudoscalar mixing in $J/\psi$ and $\psi(2S)$
decays has been analyzed, which shows $\eta$ favors only consist of
light quarks and $\eta'$ has a room for gluonium admixture. However,
in this paper we shall only consider their quark content in order to
analyze the properties of coupling constants from $SU(3)$ breaking.
The results of fit are given in Table 3. From this table, we can
obtain the following results: (i) the coupling constant of the octet
strong interaction, $g_8$, is about twice larger than that one of
the singlet, $g_1$; (ii) the  $SU(3)$ breaking coupling constant
from electromagnetic effect is large, about the same order of $g_8$
and $g_1$, however, the $SU(3)$ breaking coupling constant from mass
effect is rather small, about smaller one order than those of $g_8$
and $g_1$, moreover its uncertainty is also very large; (iii) the
phase angle between strong and electromagnetic interaction is about
$\frac{2}{5}\pi$; (iv) the mixing angle in $\eta$ and $\eta'$,
$\theta_P$, is about $-20^{\circ}$, which is consistent with the
reasonable range $-20^{\circ} \sim -10^{\circ}$ \cite{feldmann2000};
and (v) compared with the results of MarkIII and DM2 data, the goodness
of fit to PDG2012 data is very large.

\begin{table*}[htpb]
\begin{center}
{\small Table\ 2\ \  Branching fractions of $J/\psi \to VP$ from MarkIII, DM2, BES and PDG2012 ($\times 10^{-3})$.}
\end{center}
\footnotesize
\begin{center}
\begin{tabular}{cccccc} \hline
Decay modes    & MarkIII                  & DM2                        & BES                        &  PDG2012        \\ \hline
$\rho\pi$    & $14.2\pm 0.1\pm 1.9$     & $13.2\pm2.0$               & $21.8\pm0.05\pm 2.01$      & $16.9\pm1.5$    \\
$\rho\eta$   & $0.193\pm0.013\pm0.029$  & $0.194\pm0.017\pm0.029$    &                            & $0.193\pm0.023$ \\
$\rho \eta'$ & $0.114\pm0.014\pm0.016$  & $0.083\pm0.030\pm0.012$    &                            & $0.105\pm0.018$ \\
$\phi\pi $   & $<0.0068$                &                            & $<0.0064$                  & $<0.0064$       \\
$\phi\eta$   & $0.661\pm0.045\pm0.078$  & $0.64\pm0.04\pm0.11$       & $0.898\pm0.024\pm0.089$    & $0.75\pm0.08$   \\
$\phi\eta'$   &$0.308\pm0.034\pm0.036$  & $0.41\pm0.03\pm0.08$       & $0.546\pm0.031\pm0.056$    & $0.40\pm0.07$   \\
$\omega\pi $  &$0.482\pm0.019\pm0.064$  & $0.360\pm0.028\pm0.054$    & $0.538\pm0.012\pm0.065$    & $0.45\pm0.05$   \\
$\omega\eta$  &$1.71\pm0.08\pm0.20$     & $1.43\pm0.10\pm0.21$       & $2.352\pm0.273$            & $1.74\pm0.20$   \\
$\omega\eta'$ &$0.166\pm0.017\pm0.019$  & $0.18^{+0.10}_{-0.08}\pm0.03$ & $0.226\pm0.043$         & $0.182\pm0.021$ \\
$K^{*-}K^+ + {\rm c.c.}$ &$5.26\pm0.13\pm0.53$   & $4.57\pm0.17\pm0.70$       &                            & $5.12\pm0.30$   \\
$K^{*0} \bar{K^{0}}+{\rm c.c.}$& $4.33\pm 0.12\pm 0.45$ & $3.96\pm0.15\pm0.60$ &                         & $4.39\pm0.31$   \\ \hline
\end{tabular}
\end{center}
\end{table*}

\subsection{Analysis of $J/\psi \to VP$ from MarkIII, DM2 and PDG2012 data in Case II}

\begin{table*}[htpb]
\begin{center}
{\small Table\ 3\ \  Results of fit to MarkIII, DM2 and PDG2012 data in Case I.}
\end{center}
\footnotesize
\begin{center}
\begin{tabular}{cccc}
 \hline  Parameter               &   MarkIII                & DM2                    & PDG2012        \\ \hline
 $g_8$ $(\times 10^{-3})$   &   $5.89 \pm 0.18$        & $5.57 \pm 0.22$        & $5.98 \pm 0.13$ \\
 $g_1$ $(\times 10^{-3})$   &   $2.72 \pm 0.20$        & $3.12 \pm 0.40$        & $2.77 \pm 0.26$ \\
 $g_M$ $(\times 10^{-4})$   &   $8.08 \pm 3.21$        & $5.84 \pm 4.22$        & $9.44 \pm 2.72$ \\
 $|g_E|$ $(\times 10^{-3})$ &   $2.11 \pm 0.10$        & $1.95 \pm 0.11$        & $2.06 \pm 0.08$ \\
 $\delta_E$                 &   $71.4 \pm 11.5$        & $74.9 \pm 17.0$        & $75.8 \pm 6.87$ \\
 $\theta_P$                 &   $-19.5 \pm 1.52$       & $-20.0 \pm 0.68$       & $-19.0\pm 1.59$ \\ \hline
$\chi^2/d.o.f$              &   7.19/4                 & 3.63/4                 & 19.4/4 \\ \hline
\end{tabular}
\end{center}
\end{table*}

Next, we consider case II, i.e., $\omega-\phi$ mixing angle is left
as a free parameter.  The results of fit are listed in Table 4. From
this table, we can see that, the results of this case is similarly
those of Case I only, but the results of fit to MarkIII data,
$\theta_P= -15.2 \pm 2.93$ are obvious difference with those of
Case I. The table also shows that the mixing in $\omega$ and $\phi$ is
basically an idea mixing.

\subsection{Analysis of the branching ratios of $J/\psi \to VP$ in Case III and IV }

Finally, we consider the latter two cases. There are 10 and 11 free
parameters in Case III and IV, respectively.  Because each of the
experimental results are not sufficient do a reasonable fit for 10 or 11 free parameters in these two cases, we try to figure out these free parameters by 
combining the data from the different collaborations. Such as for Case III,
the results of fit from MarkIII $+$ BES data can be written as
 \begin{eqnarray}
&&g_8 = (6.57 \pm 0.16) \times 10^{-3}, \nonumber \\
&&g_1 = (3.14 \pm 0.15) \times 10^{-3}, \nonumber \\
&&g^{88}_M = (18.1 \pm 4.01) \times 10^{-4}, \nonumber \\
&&g^{81}_M = (5.78 \pm 4.20) \times 10^{-4}, \nonumber \\
&&g^{18}_M = (3.23 \pm 0.21) \times 10^{-4}, \nonumber \\
&&|g^{88}_E| = (1.73 \pm 0.22) \times 10^{-3}, \nonumber \\
&&|g^{81}_E| = (2.74 \pm 0.20) \times 10^{-3}, \nonumber \\
&&|g^{18}_E| = (2.18 \pm 0.15) \times 10^{-3}, \nonumber \\
&&\delta_E=69.5 \pm 12.6,\ \ \theta_P=-19.0 \pm 3.44.
\end{eqnarray}
we can see that, (i) the coupling constants of strong interaction,
$g_8$ and $g_1$, are basically  consistent with those of Case I and
Case II; (ii)the difference of three coupling constants of
electromagnetic effect isn't obvious, but three coupling constants
of mass effect have rather large difference, moreover their
uncertainties are rather large; and (iii) the goodness of fit in this
case is very large, it is a possible reason that we deal together
with the data from the different collaborations.

\begin{table*}[htpb]
\begin{center}
{\small Table\ 4\ \  Results of fit to MarkIII, DM2 and PDG2012 data in Case II.}
\end{center}
\footnotesize
\begin{center}
\begin{tabular}{cccc}
 \hline  Parameter               &   MarkIII                & DM2                    & PDG2012        \\ \hline
 $g_8$ $(\times 10^{-3})$   &   $5.87 \pm 0.18$        & $5.57 \pm 0.22$        & $5.99 \pm 0.13$ \\
 $g_1$ $(\times 10^{-3})$   &   $2.84 \pm 0.21$        & $3.12 \pm 0.42$        & $2.84 \pm 0.29$ \\
 $g_M$ $(\times 10^{-4})$   &   $5.75 \pm 3.40$        & $5.88 \pm 4.84$        & $8.42 \pm 3.32$ \\
 $|g_E|$ $(\times 10^{-3})$ &   $2.15 \pm 0.10$        & $1.94 \pm 0.12$        & $2.07 \pm 0.08$ \\
 $\delta_E$                 &   $71.7 \pm 11.1$        & $74.9 \pm 16.9$        & $76.0 \pm 6.81$ \\
 $\theta_P$                 &   $-15.2 \pm 2.93$       & $-20.1 \pm 5.06$       & $-17.8\pm 2.68$ \\
 $\theta_V$                 &   $41.7 \pm 3.65$        & $35.4  \pm 7.27$       & $37.5 \pm 3.42$ \\\hline
$\chi^2/d.o.f$              &   4.95/3                 & 3.63/3                 & 19.2/3 \\ \hline
\end{tabular}
\end{center}
\end{table*}

For Case IV, the results of fit from MarkIII and BES data are
 \begin{eqnarray}
&&g_8 = (6.57 \pm 0.16) \times 10^{-3}, \nonumber \\
&&g_1 = (3.05 \pm 0.15) \times 10^{-3}, \nonumber \\
&&g^{88}_M = (18.7 \pm 3.30) \times 10^{-4}, \nonumber \\
&&g^{81}_M = (11.0 \pm 1.47) \times 10^{-4}, \nonumber \\
&&g^{18}_M = (7.00 \pm 0.68) \times 10^{-4}, \nonumber \\
&&|g^{88}_E| = (1.81 \pm 0.22) \times 10^{-3}, \nonumber \\
&&|g^{81}_E| = (2.70 \pm 0.20) \times 10^{-3}, \nonumber \\
&&|g^{18}_E| = (2.14 \pm 0.15) \times 10^{-3}, \nonumber \\
&&\delta_E=78.0 \pm 2.56,\ \ \theta_P=-18.0 \pm 0.53, \nonumber \\
&&\theta_V=35.9 \pm 1.47, \ \ \chi^2/d.o.f=36.1/5.
\end{eqnarray}
This case is similar to Case III, the difference of three coupling
constants of  electromagnetic effect isn't still obvious, three
coupling constants of mass effect have rather large difference, and
at the same time, the phase angle between strong and electromagnetic
interaction is slightly large.

The results of fit from DM2$+$BES combination and MarkIII $+$ DM2
combination, are similar to those of MarkIII $+$ BES. Their numerical
results aren't individually given out.

\section{Conclusions}

Basing on the general phenomenological model, we have studied the
properties  of the coupling constants of  the decays $J/\psi \to
VP$. Considering the experimental data of MarkIII, DM2 and BES
collaborations and the world average in 2012 for $J/\psi \to VP$
decays, we can find that (i) the octet coupling constant of strong
interaction $g_8$ is about twice larger than that of the singlet
coupling constant $g_1$; (ii) the electromagnetic breaking terms
$g^i_E$ are larger, about the same order of $g_8$ and $g_1$, the
difference of three coupling constants $g^i_E$ isn't obviously, then
it is appropriate that we take them regard as equal in Case I and
II; (iii) the $SU(3)$ breaking coupling constant from mass effect is
rather small, about one order smaller than those of $g_8$ and $g_1$.
Moreover its uncertainty is quite large. It is a very strong
assumption that we take them regard as equal in Case I and II; (iv)
the phase angle between strong and electromagnetic interaction is
about the range of $70^{\circ} \sim 80^{\circ}$; (v) the mixing
angle in $\eta$ and $\eta'$, $\theta_P$, is about the range of
$-15^{\circ}- -20^{\circ}$, which is consistent with the
reasonable range usually considered; (vi) the mixing angle
$\theta_V$ is about the rang of $35^{\circ} \sim 37^{\circ}$, it
means the $\phi-\omega$ mixing is basically an idea mixing.
Therefore in the breaking of $SU(3)$ favor symmetry, the
electromagnetic effects are large, moreover the mass effects are
comparatively small, which affords useful information to comprehend
the breaking of $SU(3)$ flavor symmetry.

\vspace{-3mm}

\section*{ACKNOWLEDGMENTS}

This project was supported in part by the National Natural Science
Foundation of China under Grant Nos. 10979008, 11275957 and U1232101; the
Program for Science and Technology Innovation Talents in Universities of He'nan Province under No. 2012HASTIT030.

\end{document}